\documentclass[aps,prb,superscriptaddress,twocolumn]{revtex4-1}% Physical Review B

\usepackage{bm, amssymb}
\usepackage{dcolumn}
\usepackage{graphicx}
\usepackage[dvips]{epsfig}
\usepackage{xcolor}
\usepackage{bm, amssymb}
\usepackage{dcolumn}

\begin{document}
	
	\title{Origin of the high Seebeck coefficient of the misfit [Ca$_2$CoO$_3$]$_{0.62}$[CoO$_2$] cobaltate from site-specific valency and spin-state determinations}
	\author{Abdul Ahad}
	\address{Department of Physics, Aligarh Muslim University, Aligarh 202002, India}
	\author{K. Gautam}
	\address{UGC-DAE Consortium for Scientific Research, Indore 452001, India}
	\author{S. S. Majid}
	\thanks{Present address: Optical Physics Lab, Institute of Physics, Academia Sinica, Taipei, Taiwan}
	\address{Department of Physics, Aligarh Muslim University, Aligarh 202002, India}
	\author{S. Francoual}
	\address{Deutsches Elektronen-Synchrotron, Notkestrasse 85, D-22607 Hamburg, Germany}
	\author{F. Rahman}
	\address{Department of Physics, Aligarh Muslim University, Aligarh 202002, India}
	\author{Frank M. F. De Groot}
	\address{Debye Institute for Nanomaterials Science, Utrecht University-99, 3584 CG, Utrecht, The Netherlands}
	\author{D. K. Shukla}
	\thanks{Corresponding Author: dkshukla@csr.res.in}
	\address{UGC-DAE Consortium for Scientific Research, Indore 452001, India}

\date{\today}

\begin{abstract}
Layered misfit cobaltate [Ca$_2$CoO$_3$]$_{0.62}$[CoO$_2$], which emerged as an important thermoelectric material~[A. C. Masset $et~al.$ Phys. Rev. B, 62, 166 (2000)], has been explored extensively in the last decade for the exact mechanism behind its high Seebeck coefficient. Its complex crystal and electronic structures have inhibited consensus among such investigations. This situation has arisen mainly due to difficulties in accurate identification of the chemical state, spin state, and site symmetries in its two subsystems (rocksalt [Ca$_2$CoO$_3$] and triangular [CoO$_2$]). By employing resonant photoemission spectroscopy and x-ray absorption spectroscopy along with charge transfer multiplet simulations (at the Co ions), we have successfully identified the site symmetries, valencies and spin states of the Co in both layers. Our site-symmetry observations explain the experimental value of the high Seebeck coefficient and also confirm that the carriers hop within the rocksalt layer, which is in contrast to earlier reports where hopping within triangular CoO$_2$ layer has been held responsible for the large Seebeck coefficient.
\end{abstract}

\maketitle

Materials that can convert heat into electricity are often called as thermoelectric materials. A good thermoelectric material should possess low thermal conductivity ($\kappa$), high Seebeck coefficient (S), and high electrical conductivity ($\sigma$) to provide maximum value of figure of merit ZT (S$^2$ $\sigma$/$\kappa$). The cobalt-based layered structure family (Na$_x$CoO$_2$, Bi$_2$Sr$_2$Co$_2$O$_9$ and Ca$_3$Co$_4$O$_9$) fulfilling the above-mentioned requirements has become popular. Especially the cobaltates, Na$_x$CoO$_2$~\cite{Tak2003} and Bi$_2$Sr$_2$Co$_2$O$_9$~\cite{Kan2006}, with the triangular CoO$_2$ lattice which is made up of edge shared trigonal symmetric CoO$_6$ octahedra have gathered much attraction. The former is very well studied because of its unique properties like superconductivity in the hydrated form~\cite{Tak2003,Sch2003} and high S value for x = 0.5 composition~\cite{Kos2000}. The later one also has a high S value and its properties were explored using several techniques such as photoemission and absorption spectroscopies~\cite{Kan2006}. The discovery of large S in Na$_x$CoO$_2$ opened a path for researchers to make efforts in these structures. The chemical stability of the thermoelectric materials at high temperatures is also a common issue from the application point of view and this restricts the use of Na.

Earlier, Ca$_3$Co$_4$O$_9$ (CCO) emerged as an important candidate for thermoelectricity from the misfit cobaltate family with stability up to high temperatures~\cite{Mas2000}. Its crystal structure comprises two incommensurately modulated subsystems sharing the same $a$ and $c$ but different $b$ lattice parameters (for details, see the Supplemental Material~\cite{Sup} and references~\cite{Miy2002,Sug2002} therein). The chemical structure of CCO (precise chemical formula [Ca$_2$CoO$_3$]$_{0.62}$[CoO$_2$]) is similar to that of Na$_x$CoO$_2$ and can be compared with its $x$=2/3 composition. In CCO, the layer CoO$_2$ is presumed to be conducting and the rocksalt layer Ca$_2$CoO$_3$ insulating, as is suggested from studies on other iso-structural compounds~\cite{Kan2006,Miz2005}. Mixed valency of Co$^{3+}$ and Co$^{4+}$ is reported by x-ray photoemission and absorption spectroscopy~\cite{Wak2008,Miz2005} and the Heikes formula was employed to calculate S values which was first used by Koshibae $et~al.$~\cite{Kos2000} for explaining the S value, in high temperature, for mixed valent cobaltates. However, the origin of large S in this compound is controversial and has been proposed in different ways. By electron energy loss spectroscopy measurements, it was shown that to maintain charge neutrality, holes from the rocksalt layers transfer into CoO$_2$ layer and increase the concentration of mobile holes in it, which enables high S [Ref.~\onlinecite{Yan2008}]. Also, using $2p-3d$ resonant photoemission spectroscopy (RPES), it is reported~\cite{Tak2004} that the Co $3d$ and O $2p$ hybridized states are spread from E$_F$ up to 8 eV and S has been calculated using Boltzmann metallic conduction with extended band, not by the Heikes formula. Moreover, theoretically, application of the Heikes formula is reported in rocksalt~\cite{Asa2002} as well as in the CoO$_2$ layer, both,~\cite{Reb2012,Sor2012} and these controversies continue because of the lack of experimental evidence.

Note that the Co valency estimation from an approximate chemical formula (Ca$_3$Co$_4$O$_9$) gives Co in +3 oxidation state, while in the misfit form the chemical formula contains CoO$_2$ and Ca$_2$CoO$_3$, which, individually supposedly contains Co ions in +4 and +2 states, respectively. However, on comparing its actual chemical formula [Ca$_2$CoO$_3$]$_{0.62}$[CoO$_2$] with equivalent Na$_{2/3}$CoO$_2$, one gets a clue that the cobalt in CoO$_2$ layers has a +3.34 oxidation state and +3.05 in rocksalt layer. This suggests that the holes should transfer from the CoO$_2$ to the rocksalt layer but the scenario reported~\cite{Yan2008} is contrary to this. Unfortunately, no direct tool exist that can estimate the correct valency in these two subsystems. The literature also contains controversies regarding the spin states of the Co$^{3+}$ and the Co$^{4+}$ ions and also which layer is the conducting layer - rocksalt or triangular.

In this report, we address the above issues and unravel the observation of different valency and spin states in rock salt and the CoO$_2$ layers by utilizing the symmetry as a distinction tool for two subsystems. The results are unique and provide evidence of the presence of both D$_{3d}$ and D$_{4h}$ symmetries. Cobalt ions in mixed spin states [high spin state (HSS) + low spin state (LSS)] and +3/+4 valency are found to play an important role in the rocksalt layer for high Seebeck coefficient. We also confirm that density of states (DOS) of the triangular layer is adjacent to E$_F$ and contains Co in +3 only. Based on our results of chemical states and spins states in both these layers we have employed the Heikes formula in rocksalt and calculated value of S is in excellent agreement with the experimental value.

Polycrystalline CCO has been synthesized using solid-state route using postcalcination method~\cite{Kan2014}. X-ray diffraction (XRD) has been performed at P09, DESY, Germany with 0.539 \AA~wavelength using image plate detector (Perkin Elmer XRD1621 detector having 40 $\times$ 40 cm$^2$ active area with 2048 $\times$ 2048 pixels). Single phase synthesis is confirmed by Rietveld refinement using monoclinic superspace group $C2/m(0b0)00$, where $b$=1.612 is the structural modulation vector~\cite{Gre2004} (see Supplemental Material~\cite{Sup} Figs. S1 \& S2). Magnetic susceptibility and resistivity has been measured as a function of temperature to confirm the quality of the sample (see Fig. S3 (a \& b)~\cite{Sup}). X-ray photoemission spectroscopy (XPS) has been carried out using Omicron energy analyzer (EA-125) with Al $K\alpha$ (1486.6 eV) x-ray source. Valence band spectra (VBS) with the incident photon energies in the range of 44 - 68 eV were recorded at BL-02 of Indus-1 synchrotron, RRCAT, India . The experimental resolution in this photon energy range was estimated to be $\sim$0.3 - 0.4 eV. The X-ray absorption spectroscopy (XAS) experiments were performed at BL-01 of Indus-2 synchrotron, RRCAT, India. In the XAS experiments, energy resolution at the Co L$_{3,2}$ edges was $\sim$0.3 eV.

\begin{figure}[hbt]
\includegraphics[width=0.49\textwidth]{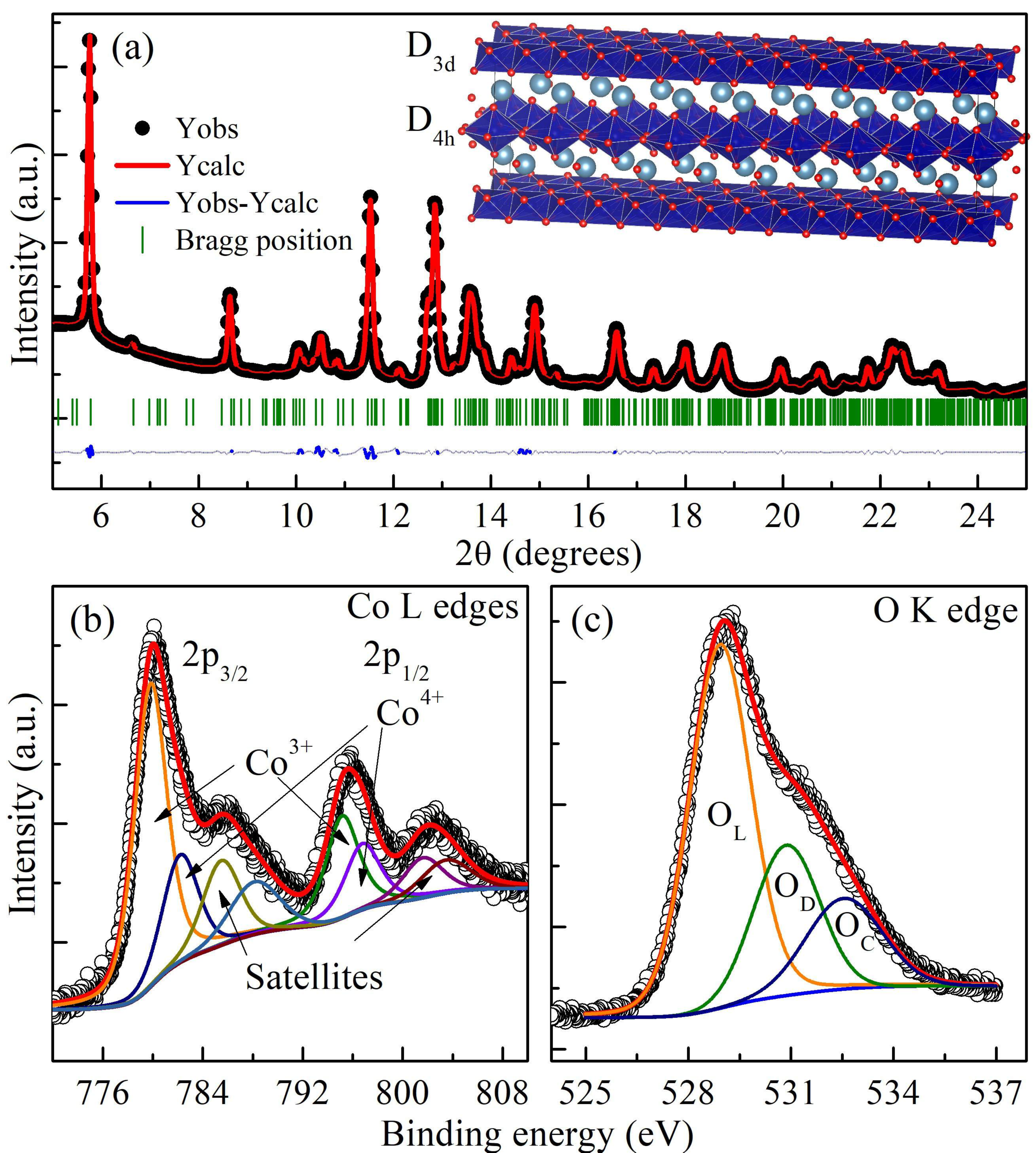}
\caption{(a) Rietveld refined XRD pattern with the supercell in inset showing cobalt sites in rocksalt and triangular layers. (b) Co 2p XPS spectra fitted with Co$^{3+}$ and Co$^{4+}$ components. (c) O 1s core XPS with lattice oxygen (O$_L$), deficient oxygen (O$_D$) and chemi-absorbed (O$_C$) oxygens features.}
\label{xrdxps}
\end{figure}

Fig.~\ref{xrdxps} (a) shows the Rietveld refined XRD pattern of CCO. Inset shows the supercell ($a, 13b, c$) comprising both subsystems (see Fig. S1 \& S2 and related text~\cite{Sup}). It shows the cobalt environments in triangular CoO$_2$ layer and rocksalt layer. Fitted Co $2p$ and O $1s$ core XPS are displayed in Figs.~\ref{xrdxps} (b) \& (c). XPS fitting reveals the +3 ($\sim$68$\%$) and +4 ($\sim$32$\%$) oxidation states of the cobalt ion in CCO. Therefore, average cobalt valency is found to be +3.32. The O $1s$ XPS shows lattice oxygen deficiency which may act as an electron doping at the cobalt sites in the rocksalt layer $i.e.$ [Ca$_2$CoO$_{3-\delta}$]$_{0.62}$[CoO$_2$]. The mixed valency can not tell which layer contains how much proportion of an oxidation state (+3 or +4) nor about the spin states.

%\begin{figure}[hbt]
%\includegraphics[width=0.47\textwidth]{xps.eps}
%\caption{(left panel) Co 2p XPS spectra fitted with Co$^{3+}$ and Co$^{4+}$ components, (right panel) O 1s core XPS with lattice oxygen (O$_L$), deficient oxygen (O$_D$) and chemi-absorbed (O$_C$) oxygens.}
%\label{xps}
%\end{figure}

\begin{figure}[hbt]
\includegraphics[width=0.48\textwidth]{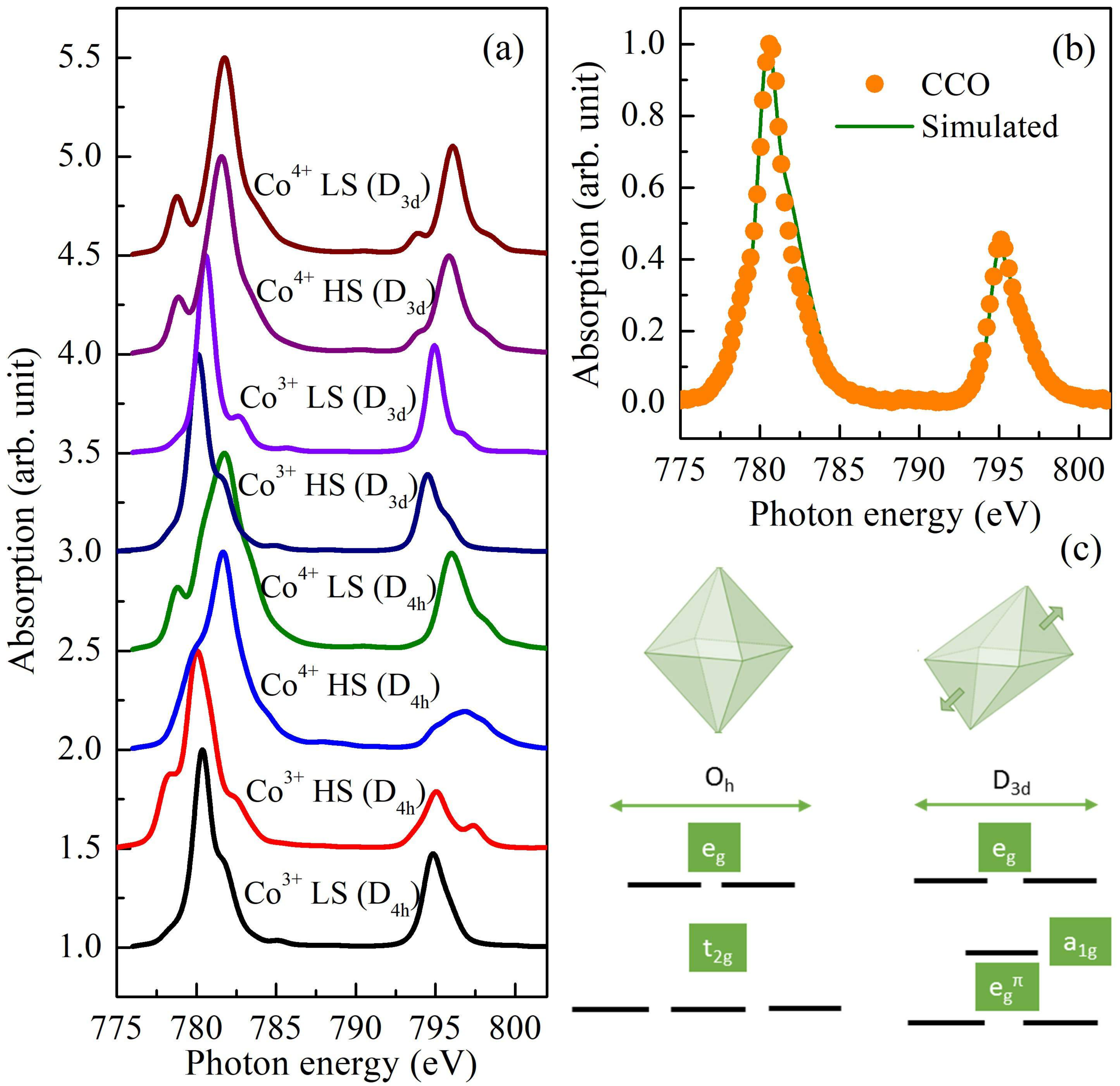}
\caption{(a) Simulated XAS spectra of Co$^{3+}$ and Co$^{4+}$ ions in HSS and LSS under D$_{4h}$ and D$_{3d}$ crystal fields. (b) Experimental and simulated XAS spectra of CCO. (c) Schematic of the crystal field effects in O$_h$ and D$_{3d}$ symmetries on degeneracy of $d$ orbitals.}
\label{xas}
\end{figure}

In literature, previous reports have shown the Co$^{3+}$ and Co$^{4+}$ in LSSs~\cite{Reb2012,Miz2005}. Generalized gradient approximation (GGA) calculations~\cite{Asa2002} concluded the Co$^{3+}$ in HSS and Co$^{4+}$ in intermediate spin state. To investigate spin-states, we performed XAS measurements. This spectroscopy is able to probe spin states which appear in the multiplet feature(s) changes as the orbital occupation changes~\cite{Abd2017} as well as symmetry of crystal field~\cite{Gro2008}. Note that the orbital occupation depends upon the local symmetry around the metal ion. Usually, symmetry of octahedra depends on the type of connectivity. For example, corner shared octahedra generally accepts high symmetry (O$_h$ or D$_{4h}$) while the edge-shared, low symmetry (D$_{3d}$). In the case of O$_h$ symmetry the resulting cubic crystal field splits the metal $d$ orbitals into $e_g$ and $t_{2g}$ orbitals which further splits into $a_{1g}$, $b_{1g}$ and $e^\pi_g$, $b_{2g}$, in lower symmetry like D$_{4h}$ (tetragonal crystal field). For the D$_{3d}$ symmetry (trigonal crystal field) case, octahedra is compressed along the (111)-direction~\cite{Sor2012}, and degeneracy in $e_g$ orbitals exists but $t_{2g}$ splits into $a_{1g}$ and $e_g^\pi$. It is known that Co environment in CoO$_2$ layer is in the D$_{3d}$ symmetry and the rocksalt layers possess octahedra with distorted O$_h$ symmetry~\cite{Pra2013}. Therefore, for the calculation of XAS patterns using the charge transfer multiplet simulation~\cite{Sta2010}, we have considered the symmetry in the rocksalt as D$_{4h}$. Here we used the hopping parameters as $T_{eg}$ = 2 eV \& $T_{t2g}$ = 1 eV and, to include the hybridization between states we reduced the atomic multiplet to 80\% ($i.e.$ Slater integral $F_{dd}$=$F_{pd}$=$G_{pd}$=0.8). Distortion parameters D$_s$ \& D$_t$ (for D$_{4h}$) and D$_{\sigma}$ \& D$_{\tau}$ (for D$_{3d}$) have been calculated from $\Delta e_g$ \& $\Delta t_{2g}$ using relations reported elsewhere~\cite{Gro2008,Hav2005}. Other parameters in the simulation are adopted from literature~\cite{Mer2011,Lin2010} and tabulated in Table S1~\cite{Sup}. Fig.~\ref{xas} (a) shows simulated XAS patterns for Co$^{3+}$ and Co$^{4+}$ in LSS and HSS in D$_{4h}$ as well as D$_{3d}$ crystal field symmetry. Fig.~\ref{xas} (b) shows the experimentally observed XAS spectra and a simulated XAS spectra which is an iterative mixing of patterns shown in (a). Combination of Co$^{3+}$/Co$^{4+}$ valencies and their spin states under D$_{4h}$ and D$_{3d}$ symmetries which resulted the best fit (Fig.~\ref{xas} (b)) are tabulated in Table~\ref{fit}. Note that this combination is obtained under the constraints that fractions of Co$^{3+}$ and Co$^{4+}$ are 68\% and 32\%, respectively, as observed from XPS.

\begin{table}[t] %add [H] placement to break table across pages
\caption{Concentration of Co ions with different spin states and valencies in D$_{4h}$ and D$_{3d}$ symmetry.}
\centering
\label{fit}
\begin{ruledtabular} %start with double line
\begin{tabular}{c c c}
Ion & D$_{4h}$ & D$_{3d}$\\
\hline
Co$^{3+}$ HSS & 20\% & 0\\
Co$^{3+}$ LSS & 14\% & 34\%\\
Co$^{4+}$ HSS & 0 & 0\\
Co$^{4+}$ LSS & 32\% & 0\\
\end{tabular}
\end{ruledtabular}
\end{table}

\begin{figure}[hbt]
\includegraphics[width=0.49\textwidth]{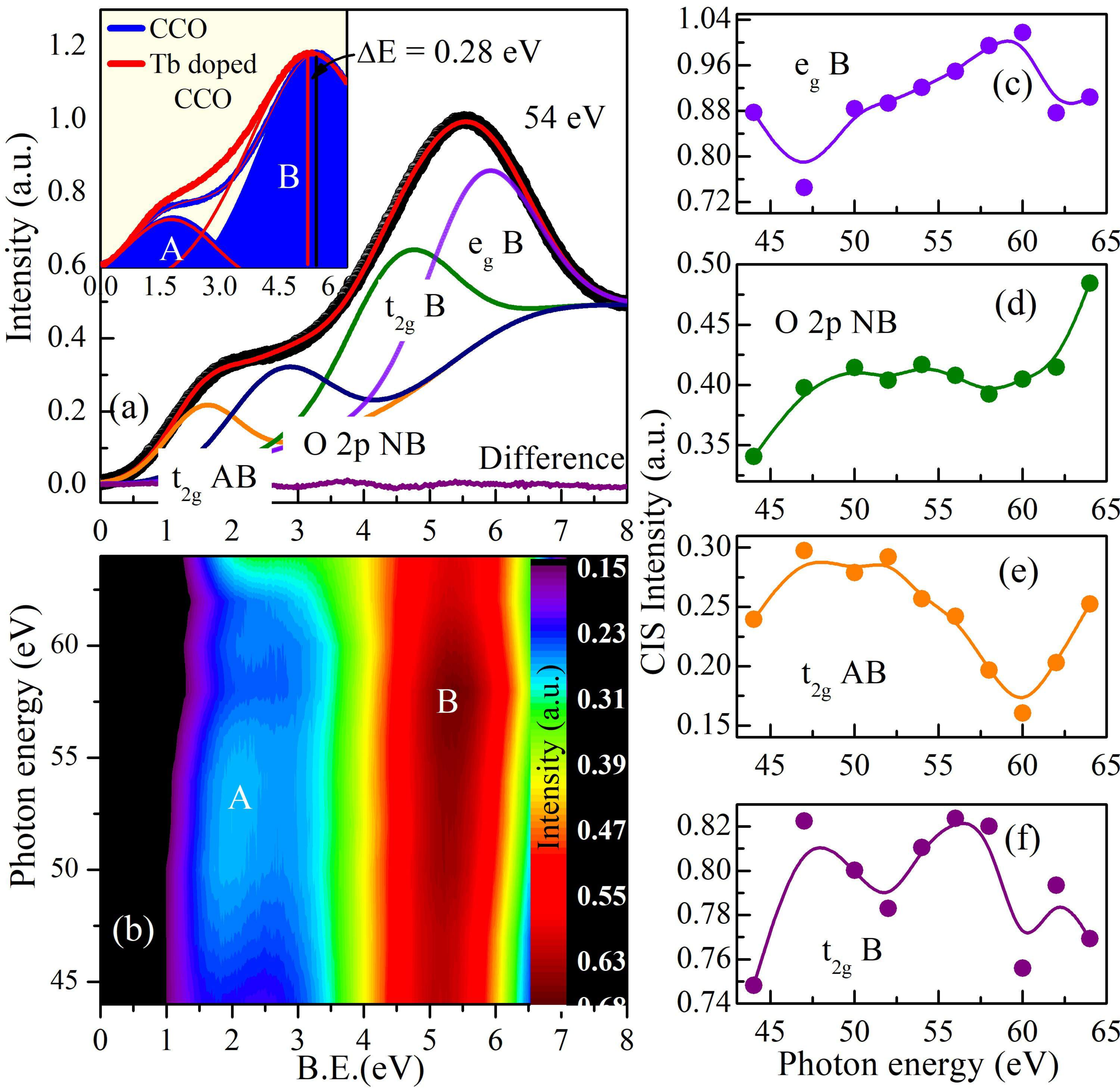}
\caption{(a) Fitted VBS with four bands of O$_h$ symmetry [Ref.~\onlinecite{Tak2004}], oscillations in the difference curve (fitted - measured) shows the inconsistency of the model used. Inset shows the VBS of CCO and Tb-doped CCO, measured at 44 eV. (b) Contour plot of VBS measured at different energies (48-68 eV). (c)-(f) CIS plots for particular features in VBS.}
\label{rpesold}
\end{figure}

Fig.~\ref{rpesold} shows the RPES results, the valence band spectroscopy in the $3p$-$3d$ resonance region (44-68 eV). In this energy interval, there may be two favorable excitations: first, direct photoemission and, second, super Coster-Kronig decay, which are given as $3p^63d^n+h\nu \longrightarrow 3p^63d^{n-1}+e^-$ and $3p^63d^n+h\nu \longrightarrow [3p^53d^{n+1}]^*\longrightarrow 3p^63d^{n-1}+e^-$, respectively, and the interference between these two give rise to resonance~\cite{Dav1986} in the intensity of $3d$ dominated bands in the valence band. Using the results of reported~\cite{Tak2004} $2p$-$3d$ RPES, we fitted the VBS using four peaks as $t_{2g}$ anti-bonding (AB), $O~2p$ non-bonding (NB), $t_{2g}$ bonding (B) and $e_g$ bonding (B). These assignments are made by assuming the O$_h$ crystal field. Inadequate fitting (Fig.~\ref{rpesold} (a)) reveals the failure of this model. The constant initial state (CIS) plots show resonance and anti-resonance features. Although these resonance and anti-resonance are poorly visible, yet give hints of 3d dominance. Moreover, the contour plot (see Fig.~\ref{rpesold} (b)) clearly shows the two resonances at $\sim$52.5 eV (feature A) and $\sim$58 eV (feature B). These resonances have not been observed before and are indispensable to symmetry related information.

\begin{figure}[hbt]
\includegraphics[width=0.49\textwidth]{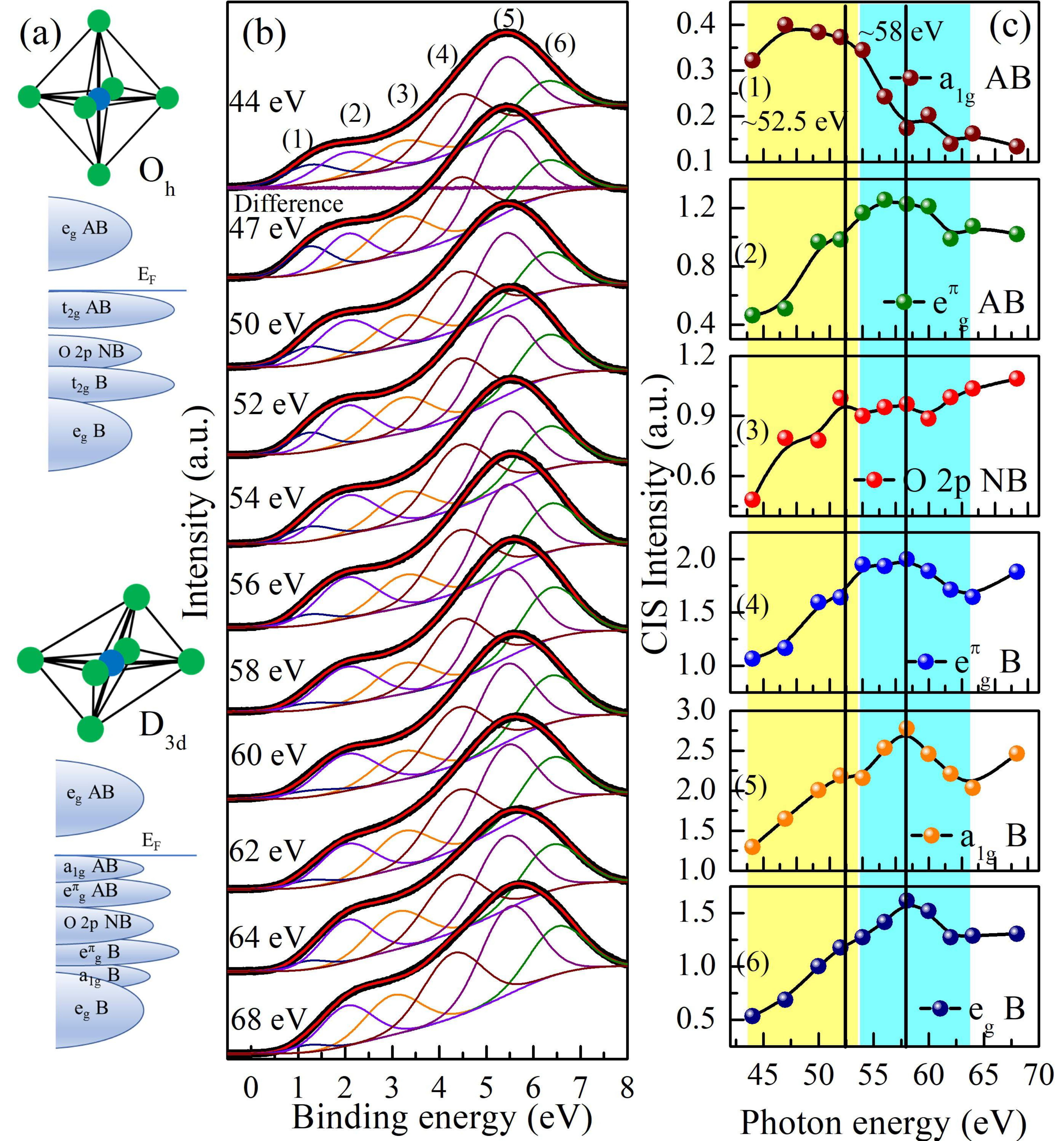}
\caption{(a) Schematic of regular octahedra in O$_h$ symmetry and its effect on degeneracy of molecular orbitals and octahedra in D$_{3d}$ symmetry and its associated molecular orbitals. (b) Fitted VBS spectra measured with different energies (across $3p$-$3d$ resonance), straight line behavior of difference curve (fitted-measured) shows the consistency of the model used and (c) CIS plots for particular features in VBS.}
\label{rpesnew}
\end{figure}

We propose that the valence band be defined by combination of D$_{4h}$ and D$_{3d}$ crystal fields as is observed from the XAS. But separation of these (D$_{4h}$ and D$_{3d}$) in RPES is not feasible due to the resolution limitations. However, to include D$_{3d}$ symmetry, the total number of participating orbitals must increase, as shown in Fig.~\ref{rpesnew} (a). We have assigned six features as $a_{1g}$ AB, $e_g^\pi$ AB, $O~2p$ NB, $e_g^\pi$ B, $a_{1g}$ B and $e_g$ B to define the VBS. In this model the CIS plots (Fig.~\ref{rpesnew} (c)) clearly show two resonances (corresponding to feature A \& B) energies in $3d$ bands.

In the earlier reported model for the CoO$_2$ layer with O$_h$ symmetry, the $t_{2g} AB$ band is near E$_F$. However, in the present modified scheme of bands, the main contribution near E$_F$ is from the $a_{1g}$ AB band as shown in the CIS plot. Its resonance for photon energy $\sim$52.5 eV confirms that the feature A which corresponds to the triangular CoO$_2$ layer has dominant Co $3d$ character. Our observation is in agreement with the $ab~initio$ theory results~\cite{Reb2012} that the CoO$_2$ layer contribute DOS near Fermi level. Resonance feature B, at a higher photon energy ($\sim$ 58 eV) is coming from the rocksalt layer. To confirm this, we have also recorded the VBS of Tb-doped CCO (Ca$_{2.9}$Tb$_{0.1}$Co$_4$O$_9$). Tb doping at the Ca site will change the Co$^{4+}$ into Co$^{3+}$. If it happens in the rocksalt layer, then this should result in the feature B moving towards E$_F$ by half of the crystal field difference between Co$^{4+}$ (10Dq $\sim$ 2.4 eV) and Co$^{3+}$ (10Dq $\sim$ 1.9 eV), which is $\sim$ 0.25 eV. Interestingly, we observe a shift of $\sim$ 0.28 eV, in feature B of Tb-doped CCO while feature A remains unchanged (see inset of Fig.~\ref{rpesold} (a) to clearly visualize the shift in the positions; both the spectra are fitted by a combination of two peaks). This is direct evidence that feature B is coming from the rocksalt layer. This ultimately confirms the band gap existence for the rocksalt subsystem as reported also by computational studies~\cite{Reb2012,Sor2012}. 

Note that earlier observation of the CoO$_2$ layer forming bands near E$_F$ along with metal-like conduction in the $ab$ plane of CCO~\cite{Mas2000} intuitively invited the proposals of S calculations based on Boltzmann metallic conduction model~\cite{Zim2001}. However, this idea has not been found truly applicable by many authors~\cite{Sor2012,Kli2012} for the reason that the temperature dependence of high S in high temperature region is not as metal, rather flat (temperature independent). Moreover, the rocksalt layer offers a band gap and, for the band insulator, S $\propto \left(E_c-~\mu \right)$ should contribute to the huge S with the decreasing trend with temperature $\left(S~\propto~\frac{1}{T}\right)$, which is also not the case~\cite{Zha2006}. Aforementioned counterintuitive scenarios motivated researchers to use the Heikes formula for understanding the origin of the high Seebeck coefficient at higher temperatures.
 
In literature, many controversies exist regarding the band positioning of particular layers and S evaluation via the Heikes formula. Asahi $et~al.$~\cite{Asa2002} using first principal GGA showed that CoO$_2$ states lie in the gap while the rocksalt contributes at the $E_F$. They calculated S in the rocksalt layer, which was 41 $\mu V$/K. Including correlation (DFT+U), Rebola $et~al.$~\cite{Reb2012} found that CoO$_2$ is actually contributing at the $E_F$ and rocksalt forms a gap and they calculated S in the CoO$_2$ layer to be 227 $\mu V$/K. Soret and Lepetit~\cite{Sor2012} using cluster quantum chemical methods with correlation, supported the results of Rebola but estimated S as 125 $\mu V$/K in the CoO$_2$ layer using non degenerate character of $a_{1g}$ orbitals. Our results related to electronic structures are consistent with these recent theoretical results but observation of the  mixed valency in rocksalt layer motivates us to utilize the Heikes formula in rocksalt layer.

According to the Heikes formula, in the high temperature limit, the thermopower can be written as~\cite{Cha1976} $S = \frac{- k_B}{e} \frac{\partial ln~g}{\partial N}$, here $k_B$ is the Boltzmann constant and $e$ is the charge. Negative sign is because of electron's negative charge. $N$ represents number of electrons and $g$ represents total number of configurations. Chaikin and Beni~\cite{Cha1976} have reported that spin degeneracy also plays an important role in determining the correct value of S. Afterward, Koshibae $et~al.$~\cite{Kos2000} introduced the factor $g_3/g_4$, ratio of the degeneracy for different valencies (+3 and +4) to further improve the approximation to the S value. The modified formula is given by Eq.~(\ref{kos})

\begin{equation}
\centering
S = \frac{- k_B}{e} ln \left( \frac{g_3}{g_4} \frac{\eta}{1 - \eta} \right)
\label{kos}
\end{equation}

%\begin{figure}[hbt]
%\includegraphics[width=0.47\textwidth]{model22.eps}
%\caption{(a) Schematics of three different states of Co$^{3+}$ and Co$^{4+}$ with the calculated spin degeneracy and orbital degeneracy g$_l$ and g$_s$, respectively. (b) Schematic of hopping of charge carriers between Co ions in rocksalt and triangular layer}
%\label{model}
%end{figure}

where $\eta$ represents the fraction of holes in the whole system. From the spectroscopic investigations, we have 14\% Co$^{3+}$ LSS, 20\% Co$^{3+}$ HSS, and 32\% Co$^{4+}$ LSS in the rocksalt layer (see Table~\ref{fit}), which we denote as $a$, $b$ and $c$, respectively. Since these make total (100\%) Co in rocksalt layer, the fractional concentration of Co$^{4+}$ LSS can be represented as $x = c/(a+b+c)$ = 0.488, Co$^{3+}$ LSS as $y = a/(a+b+c)$ = 0.21 and Co$^{3+}$ HSS as $z = b/(a+b+c)$ = 0.3. Utilizing Eq.(\ref{kos}) for two independent systems (Co$^{3+}$ HSS - Co$^{4+}$ LSS \& Co$^{3+}$ LSS - Co$^{4+}$ LSS) and assuming that probability of hopping is equal for both the Co$^{3+}$ sites one, may use $\eta = x/2 = 0.242$. Using two sites model~\cite{Spa2018} and the above configuration, one can calculate the Seebeck coefficient as, $S=\frac{-k_B}{e}\left(\left(\frac{y}{y+z}\right)ln\left(\frac{ g_{3,LSS}}{g_{4,LSS}}\frac{\eta}{1-\eta}\right) + \left(\frac{z}{y+z}\right)ln\left(\frac{g_{3,HSS}}{g_{4,LSS}} \frac{\eta}{1-\eta}\right)\right)$, which results in S $\sim$ 115.2 $\mu V/K$. This value of S is in excellent agreement and closest to the experimental value~\cite{Lim2005,Zha2006} and validates our findings.

 In conclusion, we have identified the symmetries around the Co ions in both the subsystems, triangular and rocksalt, and quantified the spin states and valencies in each subsystem. RPES results manifest the existence of two Co sites in different environments by showing resonances from each subsystem. The calculated value of S, using the Heikes formula by including the obtained spin degeneracy, $\sim$ 115.2 $\mu V/K$ is in excellent agreement with the experiments. Our results confirm that the rocksalt layer is the main contributor to the high Seebeck coefficient value of this compound. Our experiments and results not only solve the pending and debated issue of origin of temperature-independent high Seebeck coefficient of this complex misfit [Ca$_2$CoO$_3$]$_{0.62}$[CoO$_2$] cobaltate; also pave a way for spectroscopic solutions to complex compounds with non-degenerate sites and valencies.
 
%\acknowledgments
The authors thankfully acknowledge R. J. Choudhary for providing PES facility. Sharad Karwal and Rakesh Sah are acknowledged for their help during XPS/RPES and XAS measurements, respectively. The authors are grateful to Vaclav Petricek and Sander van Smaalen for stimulating discussions related to aperiodic structures. AA, KG and DKS acknowledge DST-DESY project for financial support for performing experiments at PETRA III synchrotron. AA acknowledges UGC, New Delhi, India for providing financial support in the form of the MANF scheme (2016-17/MANF-2015-17-UTT-53853). DKS acknowledges financial support from DST-New Delhi, India through INT/RUS/RFBR/P-269.

\bigskip
\bibliography{rpes_heikes}

\begin{thebibliography}{31}%
\makeatletter
\providecommand \@ifxundefined [1]{%
 \@ifx{#1\undefined}
}%
\providecommand \@ifnum [1]{%
 \ifnum #1\expandafter \@firstoftwo
 \else \expandafter \@secondoftwo
 \fi
}%
\providecommand \@ifx [1]{%
 \ifx #1\expandafter \@firstoftwo
 \else \expandafter \@secondoftwo
 \fi
}%
\providecommand \natexlab [1]{#1}%
\providecommand \enquote  [1]{``#1''}%
\providecommand \bibnamefont  [1]{#1}%
\providecommand \bibfnamefont [1]{#1}%
\providecommand \citenamefont [1]{#1}%
\providecommand \href@noop [0]{\@secondoftwo}%
\providecommand \href [0]{\begingroup \@sanitize@url \@href}%
\providecommand \@href[1]{\@@startlink{#1}\@@href}%
\providecommand \@@href[1]{\endgroup#1\@@endlink}%
\providecommand \@sanitize@url [0]{\catcode `\\12\catcode `\$12\catcode
  `\&12\catcode `\#12\catcode `\^12\catcode `\_12\catcode `\%12\relax}%
\providecommand \@@startlink[1]{}%
\providecommand \@@endlink[0]{}%
\providecommand \url  [0]{\begingroup\@sanitize@url \@url }%
\providecommand \@url [1]{\endgroup\@href {#1}{\urlprefix }}%
\providecommand \urlprefix  [0]{URL }%
\providecommand \Eprint [0]{\href }%
\providecommand \doibase [0]{http://dx.doi.org/}%
\providecommand \selectlanguage [0]{\@gobble}%
\providecommand \bibinfo  [0]{\@secondoftwo}%
\providecommand \bibfield  [0]{\@secondoftwo}%
\providecommand \translation [1]{[#1]}%
\providecommand \BibitemOpen [0]{}%
\providecommand \bibitemStop [0]{}%
\providecommand \bibitemNoStop [0]{.\EOS\space}%
\providecommand \EOS [0]{\spacefactor3000\relax}%
\providecommand \BibitemShut  [1]{\csname bibitem#1\endcsname}%
\let\auto@bib@innerbib\@empty
%</preamble>
\bibitem [{\citenamefont {Takada}\ \emph {et~al.}(2003)\citenamefont {Takada},
  \citenamefont {Sakurai}, \citenamefont {Takayama-Muromachi}, \citenamefont
  {Izumi}, \citenamefont {Dilanian},\ and\ \citenamefont {Sasaki}}]{Tak2003}%
  \BibitemOpen
  \bibfield  {author} {\bibinfo {author} {\bibfnamefont {K.}~\bibnamefont
  {Takada}}, \bibinfo {author} {\bibfnamefont {H.}~\bibnamefont {Sakurai}},
  \bibinfo {author} {\bibfnamefont {E.}~\bibnamefont {Takayama-Muromachi}},
  \bibinfo {author} {\bibfnamefont {F.}~\bibnamefont {Izumi}}, \bibinfo
  {author} {\bibfnamefont {R.~A.}\ \bibnamefont {Dilanian}}, \ and\ \bibinfo
  {author} {\bibfnamefont {T.}~\bibnamefont {Sasaki}},\ }\href@noop {}
  {\bibfield  {journal} {\bibinfo  {journal} {Nature}\ }\textbf {\bibinfo
  {volume} {422}},\ \bibinfo {pages} {53} (\bibinfo {year} {2003})}\BibitemShut
  {NoStop}%
\bibitem [{\citenamefont {Kang}\ \emph {et~al.}(2006)\citenamefont {Kang},
  \citenamefont {Han}, \citenamefont {Fujii}, \citenamefont {Terasaki},
  \citenamefont {Lee}, \citenamefont {Kim}, \citenamefont {Olson},
  \citenamefont {Lee}, \citenamefont {Kim},\ and\ \citenamefont
  {Min}}]{Kan2006}%
  \BibitemOpen
  \bibfield  {author} {\bibinfo {author} {\bibfnamefont {J.-S.}\ \bibnamefont
  {Kang}}, \bibinfo {author} {\bibfnamefont {S.}~\bibnamefont {Han}}, \bibinfo
  {author} {\bibfnamefont {T.}~\bibnamefont {Fujii}}, \bibinfo {author}
  {\bibfnamefont {I.}~\bibnamefont {Terasaki}}, \bibinfo {author}
  {\bibfnamefont {S.}~\bibnamefont {Lee}}, \bibinfo {author} {\bibfnamefont
  {G.}~\bibnamefont {Kim}}, \bibinfo {author} {\bibfnamefont {C.}~\bibnamefont
  {Olson}}, \bibinfo {author} {\bibfnamefont {H.}~\bibnamefont {Lee}}, \bibinfo
  {author} {\bibfnamefont {J.-Y.}\ \bibnamefont {Kim}}, \ and\ \bibinfo
  {author} {\bibfnamefont {B.}~\bibnamefont {Min}},\ }\href@noop {} {\bibfield
  {journal} {\bibinfo  {journal} {Phys. Rev. B}\ }\textbf {\bibinfo {volume}
  {74}},\ \bibinfo {pages} {205116} (\bibinfo {year} {2006})}\BibitemShut
  {NoStop}%
\bibitem [{\citenamefont {Schaak}\ \emph {et~al.}(2003)\citenamefont {Schaak},
  \citenamefont {Klimczuk}, \citenamefont {Foo},\ and\ \citenamefont
  {Cava}}]{Sch2003}%
  \BibitemOpen
  \bibfield  {author} {\bibinfo {author} {\bibfnamefont {R.~E.}\ \bibnamefont
  {Schaak}}, \bibinfo {author} {\bibfnamefont {T.}~\bibnamefont {Klimczuk}},
  \bibinfo {author} {\bibfnamefont {M.~L.}\ \bibnamefont {Foo}}, \ and\
  \bibinfo {author} {\bibfnamefont {R.~J.}\ \bibnamefont {Cava}},\ }\href@noop
  {} {\bibfield  {journal} {\bibinfo  {journal} {Nature}\ }\textbf {\bibinfo
  {volume} {424}},\ \bibinfo {pages} {527} (\bibinfo {year}
  {2003})}\BibitemShut {NoStop}%
\bibitem [{\citenamefont {Koshibae}\ \emph {et~al.}(2000)\citenamefont
  {Koshibae}, \citenamefont {Tsutsui},\ and\ \citenamefont
  {Maekawa}}]{Kos2000}%
  \BibitemOpen
  \bibfield  {author} {\bibinfo {author} {\bibfnamefont {W.}~\bibnamefont
  {Koshibae}}, \bibinfo {author} {\bibfnamefont {K.}~\bibnamefont {Tsutsui}}, \
  and\ \bibinfo {author} {\bibfnamefont {S.}~\bibnamefont {Maekawa}},\
  }\href@noop {} {\bibfield  {journal} {\bibinfo  {journal} {Phys. Rev. B}\
  }\textbf {\bibinfo {volume} {62}},\ \bibinfo {pages} {6869} (\bibinfo {year}
  {2000})}\BibitemShut {NoStop}%
\bibitem [{\citenamefont {Masset}\ \emph {et~al.}(2000)\citenamefont {Masset},
  \citenamefont {Michel}, \citenamefont {Maignan}, \citenamefont {Hervieu},
  \citenamefont {Toulemonde}, \citenamefont {Studer}, \citenamefont {Raveau},\
  and\ \citenamefont {Hejtmanek}}]{Mas2000}%
  \BibitemOpen
  \bibfield  {author} {\bibinfo {author} {\bibfnamefont {A.}~\bibnamefont
  {Masset}}, \bibinfo {author} {\bibfnamefont {C.}~\bibnamefont {Michel}},
  \bibinfo {author} {\bibfnamefont {A.}~\bibnamefont {Maignan}}, \bibinfo
  {author} {\bibfnamefont {M.}~\bibnamefont {Hervieu}}, \bibinfo {author}
  {\bibfnamefont {O.}~\bibnamefont {Toulemonde}}, \bibinfo {author}
  {\bibfnamefont {F.}~\bibnamefont {Studer}}, \bibinfo {author} {\bibfnamefont
  {B.}~\bibnamefont {Raveau}}, \ and\ \bibinfo {author} {\bibfnamefont
  {J.}~\bibnamefont {Hejtmanek}},\ }\href@noop {} {\bibfield  {journal}
  {\bibinfo  {journal} {Phys. Rev. B}\ }\textbf {\bibinfo {volume} {62}},\
  \bibinfo {pages} {166} (\bibinfo {year} {2000})}\BibitemShut {NoStop}%
\bibitem [{Sup()}]{Sup}%
  \BibitemOpen
  \href@noop {} {\bibinfo  {journal} {See Supplemental Material. Section I
  shows details about crystal structure. Section II shows magnetic and
  electrical characterization and section III contains the table of parameters
  used for XAS simulations}\ }\BibitemShut {NoStop}%
\bibitem [{\citenamefont {Miyazaki}\ \emph {et~al.}(2002)\citenamefont
  {Miyazaki}, \citenamefont {Onoda}, \citenamefont {Oku}, \citenamefont
  {Kikuchi}, \citenamefont {Ishii}, \citenamefont {Ono}, \citenamefont
  {Morii},\ and\ \citenamefont {Kajitani}}]{Miy2002}%
  \BibitemOpen
\bibfield  {journal} {  }\bibfield  {author} {\bibinfo {author} {\bibfnamefont
  {Y.}~\bibnamefont {Miyazaki}}, \bibinfo {author} {\bibfnamefont
  {M.}~\bibnamefont {Onoda}}, \bibinfo {author} {\bibfnamefont
  {T.}~\bibnamefont {Oku}}, \bibinfo {author} {\bibfnamefont {M.}~\bibnamefont
  {Kikuchi}}, \bibinfo {author} {\bibfnamefont {Y.}~\bibnamefont {Ishii}},
  \bibinfo {author} {\bibfnamefont {Y.}~\bibnamefont {Ono}}, \bibinfo {author}
  {\bibfnamefont {Y.}~\bibnamefont {Morii}}, \ and\ \bibinfo {author}
  {\bibfnamefont {T.}~\bibnamefont {Kajitani}},\ }\href@noop {} {\bibfield
  {journal} {\bibinfo  {journal} {J. Phys. Soc. Jpn.}\ }\textbf {\bibinfo
  {volume} {71}},\ \bibinfo {pages} {491} (\bibinfo {year} {2002})}\BibitemShut
  {NoStop}%
\bibitem [{\citenamefont {Sugiyama}\ \emph {et~al.}(2002)\citenamefont
  {Sugiyama}, \citenamefont {Itahara}, \citenamefont {Tani}, \citenamefont
  {Brewer},\ and\ \citenamefont {Ansaldo}}]{Sug2002}%
  \BibitemOpen
  \bibfield  {author} {\bibinfo {author} {\bibfnamefont {J.}~\bibnamefont
  {Sugiyama}}, \bibinfo {author} {\bibfnamefont {H.}~\bibnamefont {Itahara}},
  \bibinfo {author} {\bibfnamefont {T.}~\bibnamefont {Tani}}, \bibinfo {author}
  {\bibfnamefont {J.~H.}\ \bibnamefont {Brewer}}, \ and\ \bibinfo {author}
  {\bibfnamefont {E.~J.}\ \bibnamefont {Ansaldo}},\ }\href@noop {} {\bibfield
  {journal} {\bibinfo  {journal} {Phys. Rev. B}\ }\textbf {\bibinfo {volume}
  {66}},\ \bibinfo {pages} {134413} (\bibinfo {year} {2002})}\BibitemShut
  {NoStop}%
\bibitem [{\citenamefont {Mizokawa}\ \emph {et~al.}(2005)\citenamefont
  {Mizokawa}, \citenamefont {Tjeng}, \citenamefont {Lin}, \citenamefont {Chen},
  \citenamefont {Kitawaki}, \citenamefont {Terasaki}, \citenamefont {Lambert},\
  and\ \citenamefont {Michel}}]{Miz2005}%
  \BibitemOpen
  \bibfield  {author} {\bibinfo {author} {\bibfnamefont {T.}~\bibnamefont
  {Mizokawa}}, \bibinfo {author} {\bibfnamefont {L.}~\bibnamefont {Tjeng}},
  \bibinfo {author} {\bibfnamefont {H.-J.}\ \bibnamefont {Lin}}, \bibinfo
  {author} {\bibfnamefont {C.}~\bibnamefont {Chen}}, \bibinfo {author}
  {\bibfnamefont {R.}~\bibnamefont {Kitawaki}}, \bibinfo {author}
  {\bibfnamefont {I.}~\bibnamefont {Terasaki}}, \bibinfo {author}
  {\bibfnamefont {S.}~\bibnamefont {Lambert}}, \ and\ \bibinfo {author}
  {\bibfnamefont {C.}~\bibnamefont {Michel}},\ }\href@noop {} {\bibfield
  {journal} {\bibinfo  {journal} {Phys. Rev. B}\ }\textbf {\bibinfo {volume}
  {71}},\ \bibinfo {pages} {193107} (\bibinfo {year} {2005})}\BibitemShut
  {NoStop}%
\bibitem [{\citenamefont {Wakisaka}\ \emph {et~al.}(2008)\citenamefont
  {Wakisaka}, \citenamefont {Hirata}, \citenamefont {Mizokawa}, \citenamefont
  {Suzuki}, \citenamefont {Miyazaki},\ and\ \citenamefont
  {Kajitani}}]{Wak2008}%
  \BibitemOpen
  \bibfield  {author} {\bibinfo {author} {\bibfnamefont {Y.}~\bibnamefont
  {Wakisaka}}, \bibinfo {author} {\bibfnamefont {S.}~\bibnamefont {Hirata}},
  \bibinfo {author} {\bibfnamefont {T.}~\bibnamefont {Mizokawa}}, \bibinfo
  {author} {\bibfnamefont {Y.}~\bibnamefont {Suzuki}}, \bibinfo {author}
  {\bibfnamefont {Y.}~\bibnamefont {Miyazaki}}, \ and\ \bibinfo {author}
  {\bibfnamefont {T.}~\bibnamefont {Kajitani}},\ }\href@noop {} {\bibfield
  {journal} {\bibinfo  {journal} {Phys. Rev. B}\ }\textbf {\bibinfo {volume}
  {78}},\ \bibinfo {pages} {235107} (\bibinfo {year} {2008})}\BibitemShut
  {NoStop}%
\bibitem [{\citenamefont {Yang}\ \emph {et~al.}(2008)\citenamefont {Yang},
  \citenamefont {Ramasse},\ and\ \citenamefont {Klie}}]{Yan2008}%
  \BibitemOpen
  \bibfield  {author} {\bibinfo {author} {\bibfnamefont {G.}~\bibnamefont
  {Yang}}, \bibinfo {author} {\bibfnamefont {Q.}~\bibnamefont {Ramasse}}, \
  and\ \bibinfo {author} {\bibfnamefont {R.}~\bibnamefont {Klie}},\ }\href@noop
  {} {\bibfield  {journal} {\bibinfo  {journal} {Phys. Rev. B}\ }\textbf
  {\bibinfo {volume} {78}},\ \bibinfo {pages} {153109} (\bibinfo {year}
  {2008})}\BibitemShut {NoStop}%
\bibitem [{\citenamefont {Takeuchi}\ \emph {et~al.}(2004)\citenamefont
  {Takeuchi}, \citenamefont {Kondo}, \citenamefont {Takami}, \citenamefont
  {Takahashi}, \citenamefont {Ikuta}, \citenamefont {Mizutani}, \citenamefont
  {Soda}, \citenamefont {Funahashi}, \citenamefont {Shikano}, \citenamefont
  {Mikami} \emph {et~al.}}]{Tak2004}%
  \BibitemOpen
  \bibfield  {author} {\bibinfo {author} {\bibfnamefont {T.}~\bibnamefont
  {Takeuchi}}, \bibinfo {author} {\bibfnamefont {T.}~\bibnamefont {Kondo}},
  \bibinfo {author} {\bibfnamefont {T.}~\bibnamefont {Takami}}, \bibinfo
  {author} {\bibfnamefont {H.}~\bibnamefont {Takahashi}}, \bibinfo {author}
  {\bibfnamefont {H.}~\bibnamefont {Ikuta}}, \bibinfo {author} {\bibfnamefont
  {U.}~\bibnamefont {Mizutani}}, \bibinfo {author} {\bibfnamefont
  {K.}~\bibnamefont {Soda}}, \bibinfo {author} {\bibfnamefont {R.}~\bibnamefont
  {Funahashi}}, \bibinfo {author} {\bibfnamefont {M.}~\bibnamefont {Shikano}},
  \bibinfo {author} {\bibfnamefont {M.}~\bibnamefont {Mikami}},  \emph
  {et~al.},\ }\href@noop {} {\bibfield  {journal} {\bibinfo  {journal} {Phys.
  Rev. B}\ }\textbf {\bibinfo {volume} {69}},\ \bibinfo {pages} {125410}
  (\bibinfo {year} {2004})}\BibitemShut {NoStop}%
\bibitem [{\citenamefont {Asahi}\ \emph {et~al.}(2002)\citenamefont {Asahi},
  \citenamefont {Sugiyama},\ and\ \citenamefont {Tani}}]{Asa2002}%
  \BibitemOpen
  \bibfield  {author} {\bibinfo {author} {\bibfnamefont {R.}~\bibnamefont
  {Asahi}}, \bibinfo {author} {\bibfnamefont {J.}~\bibnamefont {Sugiyama}}, \
  and\ \bibinfo {author} {\bibfnamefont {T.}~\bibnamefont {Tani}},\ }\href@noop
  {} {\bibfield  {journal} {\bibinfo  {journal} {Phys. Rev. B}\ }\textbf
  {\bibinfo {volume} {66}},\ \bibinfo {pages} {155103} (\bibinfo {year}
  {2002})}\BibitemShut {NoStop}%
\bibitem [{\citenamefont {R{\'e}bola}\ \emph {et~al.}(2012)\citenamefont
  {R{\'e}bola}, \citenamefont {Klie}, \citenamefont {Zapol},\ and\
  \citenamefont {{\"O}{\u{g}}{\"u}t}}]{Reb2012}%
  \BibitemOpen
  \bibfield  {author} {\bibinfo {author} {\bibfnamefont {A.}~\bibnamefont
  {R{\'e}bola}}, \bibinfo {author} {\bibfnamefont {R.}~\bibnamefont {Klie}},
  \bibinfo {author} {\bibfnamefont {P.}~\bibnamefont {Zapol}}, \ and\ \bibinfo
  {author} {\bibfnamefont {S.}~\bibnamefont {{\"O}{\u{g}}{\"u}t}},\ }\href@noop
  {} {\bibfield  {journal} {\bibinfo  {journal} {Phys. Rev. B}\ }\textbf
  {\bibinfo {volume} {85}},\ \bibinfo {pages} {155132} (\bibinfo {year}
  {2012})}\BibitemShut {NoStop}%
\bibitem [{\citenamefont {Soret}\ and\ \citenamefont
  {Lepetit}(2012)}]{Sor2012}%
  \BibitemOpen
  \bibfield  {author} {\bibinfo {author} {\bibfnamefont {J.}~\bibnamefont
  {Soret}}\ and\ \bibinfo {author} {\bibfnamefont {M.-B.}\ \bibnamefont
  {Lepetit}},\ }\href@noop {} {\bibfield  {journal} {\bibinfo  {journal} {Phys.
  Rev. B}\ }\textbf {\bibinfo {volume} {85}},\ \bibinfo {pages} {165145}
  (\bibinfo {year} {2012})}\BibitemShut {NoStop}%
\bibitem [{\citenamefont {Kang}\ \emph {et~al.}(2014)\citenamefont {Kang},
  \citenamefont {Cho}, \citenamefont {Kim}, \citenamefont {Nahm}, \citenamefont
  {Yoon},\ and\ \citenamefont {Kang}}]{Kan2014}%
  \BibitemOpen
  \bibfield  {author} {\bibinfo {author} {\bibfnamefont {M.-G.}\ \bibnamefont
  {Kang}}, \bibinfo {author} {\bibfnamefont {K.-H.}\ \bibnamefont {Cho}},
  \bibinfo {author} {\bibfnamefont {J.-S.}\ \bibnamefont {Kim}}, \bibinfo
  {author} {\bibfnamefont {S.}~\bibnamefont {Nahm}}, \bibinfo {author}
  {\bibfnamefont {S.-J.}\ \bibnamefont {Yoon}}, \ and\ \bibinfo {author}
  {\bibfnamefont {C.-Y.}\ \bibnamefont {Kang}},\ }\href@noop {} {\bibfield
  {journal} {\bibinfo  {journal} {Acta Mater.}\ }\textbf {\bibinfo {volume}
  {73}},\ \bibinfo {pages} {251} (\bibinfo {year} {2014})}\BibitemShut
  {NoStop}%
\bibitem [{\citenamefont {Grebille}\ \emph {et~al.}(2004)\citenamefont
  {Grebille}, \citenamefont {Lambert}, \citenamefont {Bouree},\ and\
  \citenamefont {Pet{\v{r}}{\'\i}{\v{c}}ek}}]{Gre2004}%
  \BibitemOpen
  \bibfield  {author} {\bibinfo {author} {\bibfnamefont {D.}~\bibnamefont
  {Grebille}}, \bibinfo {author} {\bibfnamefont {S.}~\bibnamefont {Lambert}},
  \bibinfo {author} {\bibfnamefont {F.}~\bibnamefont {Bouree}}, \ and\ \bibinfo
  {author} {\bibfnamefont {V.}~\bibnamefont {Pet{\v{r}}{\'\i}{\v{c}}ek}},\
  }\href@noop {} {\bibfield  {journal} {\bibinfo  {journal} {J. Appl.
  Crystallogr.}\ }\textbf {\bibinfo {volume} {37}},\ \bibinfo {pages} {823}
  (\bibinfo {year} {2004})}\BibitemShut {NoStop}%
\bibitem [{\citenamefont {Ahad}\ \emph {et~al.}(2017)\citenamefont {Ahad},
  \citenamefont {Shukla}, \citenamefont {Rahman}, \citenamefont {Majid},
  \citenamefont {Okram}, \citenamefont {Sinha}, \citenamefont {Phase} \emph
  {et~al.}}]{Abd2017}%
  \BibitemOpen
  \bibfield  {author} {\bibinfo {author} {\bibfnamefont {A.}~\bibnamefont
  {Ahad}}, \bibinfo {author} {\bibfnamefont {D.}~\bibnamefont {Shukla}},
  \bibinfo {author} {\bibfnamefont {F.}~\bibnamefont {Rahman}}, \bibinfo
  {author} {\bibfnamefont {S.}~\bibnamefont {Majid}}, \bibinfo {author}
  {\bibfnamefont {G.}~\bibnamefont {Okram}}, \bibinfo {author} {\bibfnamefont
  {A.}~\bibnamefont {Sinha}}, \bibinfo {author} {\bibfnamefont
  {D.}~\bibnamefont {Phase}},  \emph {et~al.},\ }\href@noop {} {\bibfield
  {journal} {\bibinfo  {journal} {Acta Mater.}\ }\textbf {\bibinfo {volume}
  {135}},\ \bibinfo {pages} {233} (\bibinfo {year} {2017})}\BibitemShut
  {NoStop}%
\bibitem [{\citenamefont {De~Groot}\ and\ \citenamefont
  {Kotani}(2008)}]{Gro2008}%
  \BibitemOpen
  \bibfield  {author} {\bibinfo {author} {\bibfnamefont {F.}~\bibnamefont
  {De~Groot}}\ and\ \bibinfo {author} {\bibfnamefont {A.}~\bibnamefont
  {Kotani}},\ }\href@noop {} {\emph {\bibinfo {title} {Core level spectroscopy
  of solids}}}\ (\bibinfo  {publisher} {CRC press},\ \bibinfo {year}
  {2008})\BibitemShut {NoStop}%
\bibitem [{\citenamefont {Prasoetsopha}\ \emph {et~al.}(2013)\citenamefont
  {Prasoetsopha}, \citenamefont {Pinitsoontorn}, \citenamefont {Bootchanont},
  \citenamefont {Kidkhunthod}, \citenamefont {Srepusharawoot}, \citenamefont
  {Kamwanna}, \citenamefont {Amornkitbamrung}, \citenamefont {Kurosaki},\ and\
  \citenamefont {Yamanaka}}]{Pra2013}%
  \BibitemOpen
  \bibfield  {author} {\bibinfo {author} {\bibfnamefont {N.}~\bibnamefont
  {Prasoetsopha}}, \bibinfo {author} {\bibfnamefont {S.}~\bibnamefont
  {Pinitsoontorn}}, \bibinfo {author} {\bibfnamefont {A.}~\bibnamefont
  {Bootchanont}}, \bibinfo {author} {\bibfnamefont {P.}~\bibnamefont
  {Kidkhunthod}}, \bibinfo {author} {\bibfnamefont {P.}~\bibnamefont
  {Srepusharawoot}}, \bibinfo {author} {\bibfnamefont {T.}~\bibnamefont
  {Kamwanna}}, \bibinfo {author} {\bibfnamefont {V.}~\bibnamefont
  {Amornkitbamrung}}, \bibinfo {author} {\bibfnamefont {K.}~\bibnamefont
  {Kurosaki}}, \ and\ \bibinfo {author} {\bibfnamefont {S.}~\bibnamefont
  {Yamanaka}},\ }\href@noop {} {\bibfield  {journal} {\bibinfo  {journal} {J.
  Solid State Chem.}\ }\textbf {\bibinfo {volume} {204}},\ \bibinfo {pages}
  {257} (\bibinfo {year} {2013})}\BibitemShut {NoStop}%
\bibitem [{\citenamefont {Stavitski}\ and\ \citenamefont
  {De~Groot}(2010)}]{Sta2010}%
  \BibitemOpen
  \bibfield  {author} {\bibinfo {author} {\bibfnamefont {E.}~\bibnamefont
  {Stavitski}}\ and\ \bibinfo {author} {\bibfnamefont {F.~M.}\ \bibnamefont
  {De~Groot}},\ }\href@noop {} {\bibfield  {journal} {\bibinfo  {journal}
  {Micron}\ }\textbf {\bibinfo {volume} {41}},\ \bibinfo {pages} {687}
  (\bibinfo {year} {2010})}\BibitemShut {NoStop}%
\bibitem [{\citenamefont {Haverkort}(2005)}]{Hav2005}%
  \BibitemOpen
  \bibfield  {author} {\bibinfo {author} {\bibfnamefont {M.~W.}\ \bibnamefont
  {Haverkort}},\ }\href@noop {} {\bibfield  {journal} {\bibinfo  {journal}
  {arXiv preprint cond-mat/0505214}\ } (\bibinfo {year} {2005})}\BibitemShut
  {NoStop}%
\bibitem [{\citenamefont {Merz}\ \emph {et~al.}(2011)\citenamefont {Merz},
  \citenamefont {Fuchs}, \citenamefont {Assmann}, \citenamefont {Uebe},
  \citenamefont {Lohneysen}, \citenamefont {Nagel},\ and\ \citenamefont
  {Schuppler}}]{Mer2011}%
  \BibitemOpen
  \bibfield  {author} {\bibinfo {author} {\bibfnamefont {M.}~\bibnamefont
  {Merz}}, \bibinfo {author} {\bibfnamefont {D.}~\bibnamefont {Fuchs}},
  \bibinfo {author} {\bibfnamefont {A.}~\bibnamefont {Assmann}}, \bibinfo
  {author} {\bibfnamefont {S.}~\bibnamefont {Uebe}}, \bibinfo {author}
  {\bibfnamefont {H.~V.}\ \bibnamefont {Lohneysen}}, \bibinfo {author}
  {\bibfnamefont {P.}~\bibnamefont {Nagel}}, \ and\ \bibinfo {author}
  {\bibfnamefont {S.}~\bibnamefont {Schuppler}},\ }\href@noop {} {\bibfield
  {journal} {\bibinfo  {journal} {Phys. Rev. B}\ }\textbf {\bibinfo {volume}
  {84}},\ \bibinfo {pages} {014436} (\bibinfo {year} {2011})}\BibitemShut
  {NoStop}%
\bibitem [{\citenamefont {Lin}\ \emph {et~al.}(2010)\citenamefont {Lin},
  \citenamefont {Chin}, \citenamefont {Hu}, \citenamefont {Shu}, \citenamefont
  {Chou}, \citenamefont {Ohta}, \citenamefont {Yoshimura}, \citenamefont
  {H{\'e}bert}, \citenamefont {Maignan}, \citenamefont {Tanaka} \emph
  {et~al.}}]{Lin2010}%
  \BibitemOpen
  \bibfield  {author} {\bibinfo {author} {\bibfnamefont {H.-J.}\ \bibnamefont
  {Lin}}, \bibinfo {author} {\bibfnamefont {Y.}~\bibnamefont {Chin}}, \bibinfo
  {author} {\bibfnamefont {Z.}~\bibnamefont {Hu}}, \bibinfo {author}
  {\bibfnamefont {G.}~\bibnamefont {Shu}}, \bibinfo {author} {\bibfnamefont
  {F.}~\bibnamefont {Chou}}, \bibinfo {author} {\bibfnamefont {H.}~\bibnamefont
  {Ohta}}, \bibinfo {author} {\bibfnamefont {K.}~\bibnamefont {Yoshimura}},
  \bibinfo {author} {\bibfnamefont {S.}~\bibnamefont {H{\'e}bert}}, \bibinfo
  {author} {\bibfnamefont {A.}~\bibnamefont {Maignan}}, \bibinfo {author}
  {\bibfnamefont {A.}~\bibnamefont {Tanaka}},  \emph {et~al.},\ }\href@noop {}
  {\bibfield  {journal} {\bibinfo  {journal} {Phys. Rev. B}\ }\textbf {\bibinfo
  {volume} {81}},\ \bibinfo {pages} {115138} (\bibinfo {year}
  {2010})}\BibitemShut {NoStop}%
\bibitem [{\citenamefont {Davis}(1986)}]{Dav1986}%
  \BibitemOpen
  \bibfield  {author} {\bibinfo {author} {\bibfnamefont {L.}~\bibnamefont
  {Davis}},\ }\href@noop {} {\bibfield  {journal} {\bibinfo  {journal} {J.
  Appl. Phys.}\ }\textbf {\bibinfo {volume} {59}},\ \bibinfo {pages} {R25}
  (\bibinfo {year} {1986})}\BibitemShut {NoStop}%
\bibitem [{\citenamefont {Ziman}(2001)}]{Zim2001}%
  \BibitemOpen
  \bibfield  {author} {\bibinfo {author} {\bibfnamefont {J.~M.}\ \bibnamefont
  {Ziman}},\ }\href@noop {} {\emph {\bibinfo {title} {Electrons and phonons:
  the theory of transport phenomena in solids}}}\ (\bibinfo  {publisher}
  {Oxford university press},\ \bibinfo {year} {2001})\BibitemShut {NoStop}%
\bibitem [{\citenamefont {Klie}\ \emph {et~al.}(2012)\citenamefont {Klie},
  \citenamefont {Qiao}, \citenamefont {Paulauskas}, \citenamefont {Gulec},
  \citenamefont {Rebola}, \citenamefont {{\"O}{\u{g}}{\"u}t}, \citenamefont
  {Prange}, \citenamefont {Idrobo}, \citenamefont {Pantelides}, \citenamefont
  {Kolesnik} \emph {et~al.}}]{Kli2012}%
  \BibitemOpen
  \bibfield  {author} {\bibinfo {author} {\bibfnamefont {R.~F.}\ \bibnamefont
  {Klie}}, \bibinfo {author} {\bibfnamefont {Q.}~\bibnamefont {Qiao}}, \bibinfo
  {author} {\bibfnamefont {T.}~\bibnamefont {Paulauskas}}, \bibinfo {author}
  {\bibfnamefont {A.}~\bibnamefont {Gulec}}, \bibinfo {author} {\bibfnamefont
  {A.}~\bibnamefont {Rebola}}, \bibinfo {author} {\bibfnamefont
  {S.}~\bibnamefont {{\"O}{\u{g}}{\"u}t}}, \bibinfo {author} {\bibfnamefont
  {M.~P.}\ \bibnamefont {Prange}}, \bibinfo {author} {\bibfnamefont
  {J.}~\bibnamefont {Idrobo}}, \bibinfo {author} {\bibfnamefont {S.~T.}\
  \bibnamefont {Pantelides}}, \bibinfo {author} {\bibfnamefont
  {S.}~\bibnamefont {Kolesnik}},  \emph {et~al.},\ }\href@noop {} {\bibfield
  {journal} {\bibinfo  {journal} {Phys Rev Lett}\ }\textbf {\bibinfo {volume}
  {108}},\ \bibinfo {pages} {196601} (\bibinfo {year} {2012})}\BibitemShut
  {NoStop}%
\bibitem [{\citenamefont {Zhao}\ \emph {et~al.}(2006)\citenamefont {Zhao},
  \citenamefont {Sun}, \citenamefont {Lu}, \citenamefont {Zhu},\ and\
  \citenamefont {Song}}]{Zha2006}%
  \BibitemOpen
  \bibfield  {author} {\bibinfo {author} {\bibfnamefont {B.}~\bibnamefont
  {Zhao}}, \bibinfo {author} {\bibfnamefont {Y.}~\bibnamefont {Sun}}, \bibinfo
  {author} {\bibfnamefont {W.}~\bibnamefont {Lu}}, \bibinfo {author}
  {\bibfnamefont {X.}~\bibnamefont {Zhu}}, \ and\ \bibinfo {author}
  {\bibfnamefont {W.}~\bibnamefont {Song}},\ }\href@noop {} {\bibfield
  {journal} {\bibinfo  {journal} {Phys. Rev. B}\ }\textbf {\bibinfo {volume}
  {74}},\ \bibinfo {pages} {144417} (\bibinfo {year} {2006})}\BibitemShut
  {NoStop}%
\bibitem [{\citenamefont {Chaikin}\ and\ \citenamefont {Beni}(1976)}]{Cha1976}%
  \BibitemOpen
  \bibfield  {author} {\bibinfo {author} {\bibfnamefont {P.}~\bibnamefont
  {Chaikin}}\ and\ \bibinfo {author} {\bibfnamefont {G.}~\bibnamefont {Beni}},\
  }\href@noop {} {\bibfield  {journal} {\bibinfo  {journal} {Phys. Rev. B}\
  }\textbf {\bibinfo {volume} {13}},\ \bibinfo {pages} {647} (\bibinfo {year}
  {1976})}\BibitemShut {NoStop}%
\bibitem [{\citenamefont {Sparks}\ \emph {et~al.}(2018)\citenamefont {Sparks},
  \citenamefont {Gurlo}, \citenamefont {Gaultois},\ and\ \citenamefont
  {Clarke}}]{Spa2018}%
  \BibitemOpen
  \bibfield  {author} {\bibinfo {author} {\bibfnamefont {T.~D.}\ \bibnamefont
  {Sparks}}, \bibinfo {author} {\bibfnamefont {A.}~\bibnamefont {Gurlo}},
  \bibinfo {author} {\bibfnamefont {M.~W.}\ \bibnamefont {Gaultois}}, \ and\
  \bibinfo {author} {\bibfnamefont {D.~R.}\ \bibnamefont {Clarke}},\
  }\href@noop {} {\bibfield  {journal} {\bibinfo  {journal} {Phys. Rev. B}\
  }\textbf {\bibinfo {volume} {98}},\ \bibinfo {pages} {024108} (\bibinfo
  {year} {2018})}\BibitemShut {NoStop}%
\bibitem [{\citenamefont {Limelette}\ \emph {et~al.}(2005)\citenamefont
  {Limelette}, \citenamefont {Hardy}, \citenamefont {Auban-Senzier},
  \citenamefont {J{\'e}rome}, \citenamefont {Flahaut}, \citenamefont
  {H{\'e}bert}, \citenamefont {Fr{\'e}sard}, \citenamefont {Simon},
  \citenamefont {Noudem},\ and\ \citenamefont {Maignan}}]{Lim2005}%
  \BibitemOpen
  \bibfield  {author} {\bibinfo {author} {\bibfnamefont {P.}~\bibnamefont
  {Limelette}}, \bibinfo {author} {\bibfnamefont {V.}~\bibnamefont {Hardy}},
  \bibinfo {author} {\bibfnamefont {P.}~\bibnamefont {Auban-Senzier}}, \bibinfo
  {author} {\bibfnamefont {D.}~\bibnamefont {J{\'e}rome}}, \bibinfo {author}
  {\bibfnamefont {D.}~\bibnamefont {Flahaut}}, \bibinfo {author} {\bibfnamefont
  {S.}~\bibnamefont {H{\'e}bert}}, \bibinfo {author} {\bibfnamefont
  {R.}~\bibnamefont {Fr{\'e}sard}}, \bibinfo {author} {\bibfnamefont
  {C.}~\bibnamefont {Simon}}, \bibinfo {author} {\bibfnamefont
  {J.}~\bibnamefont {Noudem}}, \ and\ \bibinfo {author} {\bibfnamefont
  {A.}~\bibnamefont {Maignan}},\ }\href@noop {} {\bibfield  {journal} {\bibinfo
   {journal} {Phys. Rev. B}\ }\textbf {\bibinfo {volume} {71}},\ \bibinfo
  {pages} {233108} (\bibinfo {year} {2005})}\BibitemShut {NoStop}%
\end{thebibliography}%


\begin{thebibliography}{3}
\expandafter\ifx\csname natexlab\endcsname\relax\def\natexlab#1{#1}\fi
\expandafter\ifx\csname bibnamefont\endcsname\relax
  \def\bibnamefont#1{#1}\fi
\expandafter\ifx\csname bibfnamefont\endcsname\relax
  \def\bibfnamefont#1{#1}\fi
\expandafter\ifx\csname citenamefont\endcsname\relax
  \def\citenamefont#1{#1}\fi
\expandafter\ifx\csname url\endcsname\relax
  \def\url#1{\texttt{#1}}\fi
\expandafter\ifx\csname urlprefix\endcsname\relax\def\urlprefix{URL }\fi
\providecommand{\bibinfo}[2]{#2}
\providecommand{\eprint}[2][]{\url{#2}}

\bibitem[{\citenamefont{Miyazaki et~al.}(2002)\citenamefont{Miyazaki, Onoda,
  Oku, Kikuchi, Ishii, Ono, Morii, and Kajitani}}]{Miy2002}
\bibinfo{author}{\bibfnamefont{Y.}~\bibnamefont{Miyazaki}},
  \bibinfo{author}{\bibfnamefont{M.}~\bibnamefont{Onoda}},
  \bibinfo{author}{\bibfnamefont{T.}~\bibnamefont{Oku}},
  \bibinfo{author}{\bibfnamefont{M.}~\bibnamefont{Kikuchi}},
  \bibinfo{author}{\bibfnamefont{Y.}~\bibnamefont{Ishii}},
  \bibinfo{author}{\bibfnamefont{Y.}~\bibnamefont{Ono}},
  \bibinfo{author}{\bibfnamefont{Y.}~\bibnamefont{Morii}}, \bibnamefont{and}
  \bibinfo{author}{\bibfnamefont{T.}~\bibnamefont{Kajitani}},
  \bibinfo{journal}{J. Phys. Soc. Jpn.} \textbf{\bibinfo{volume}{71}},
  \bibinfo{pages}{491} (\bibinfo{year}{2002}).

\bibitem[{\citenamefont{Sugiyama et~al.}(2002)\citenamefont{Sugiyama, Itahara,
  Tani, Brewer, and Ansaldo}}]{Sug2002}
\bibinfo{author}{\bibfnamefont{J.}~\bibnamefont{Sugiyama}},
  \bibinfo{author}{\bibfnamefont{H.}~\bibnamefont{Itahara}},
  \bibinfo{author}{\bibfnamefont{T.}~\bibnamefont{Tani}},
  \bibinfo{author}{\bibfnamefont{J.~H.} \bibnamefont{Brewer}},
  \bibnamefont{and} \bibinfo{author}{\bibfnamefont{E.~J.}
  \bibnamefont{Ansaldo}}, \bibinfo{journal}{Phys. Rev. B}
  \textbf{\bibinfo{volume}{66}}, \bibinfo{pages}{134413}
  (\bibinfo{year}{2002}).

\bibitem[{\citenamefont{Limelette et~al.}(2005)\citenamefont{Limelette, Hardy,
  Auban-Senzier, J{\'e}rome, Flahaut, H{\'e}bert, Fr{\'e}sard, Simon, Noudem,
  and Maignan}}]{Lim2005}
\bibinfo{author}{\bibfnamefont{P.}~\bibnamefont{Limelette}},
  \bibinfo{author}{\bibfnamefont{V.}~\bibnamefont{Hardy}},
  \bibinfo{author}{\bibfnamefont{P.}~\bibnamefont{Auban-Senzier}},
  \bibinfo{author}{\bibfnamefont{D.}~\bibnamefont{J{\'e}rome}},
  \bibinfo{author}{\bibfnamefont{D.}~\bibnamefont{Flahaut}},
  \bibinfo{author}{\bibfnamefont{S.}~\bibnamefont{H{\'e}bert}},
  \bibinfo{author}{\bibfnamefont{R.}~\bibnamefont{Fr{\'e}sard}},
  \bibinfo{author}{\bibfnamefont{C.}~\bibnamefont{Simon}},
  \bibinfo{author}{\bibfnamefont{J.}~\bibnamefont{Noudem}}, \bibnamefont{and}
  \bibinfo{author}{\bibfnamefont{A.}~\bibnamefont{Maignan}},
  \bibinfo{journal}{Phys. Rev. B} \textbf{\bibinfo{volume}{71}},
  \bibinfo{pages}{233108} (\bibinfo{year}{2005}).

\end{thebibliography}

\end{document}